\documentclass[prl,twocolumn,groupeaddress,nofootinbib]{revtex4}

\usepackage{graphicx}
\usepackage{epsfig}
\usepackage{epstopdf}
\usepackage{amssymb}
\def\psl{\mathrel{/}{\hspace{-.125in}p}}

\begin{document}

\title{
Superluminal neutrinos at the OPERA?}

\author{Robert B. Mann$^{1}$} 

\author{Utpal Sarkar$^{2,3}$}

\affiliation{
$^{1}$ Department of Physics \& Astronomy, University of Waterloo,
Waterloo, ON N2L 3G1, Canada\\
$^{2}$Physical Research Laboratory, Ahmedabad 380009, India\\
$^{3}$McDonnell Center for the Space Sciences, 
Washington University in St. Louis, MO 63112, USA}


\begin{abstract}

We argue that the recent measurement of the neutrino velocity to be
higher than the velocity of light  could be due to one or more of several mechanisms: violation
of   Lorentz invariance,  violation of the equivalence principle in the neutrino sector, or 
a form of  dark energy originating from neutrino condensates.
 This result need not undermine
special-relativistic foundational notions of causality. We
suggest different possibilities for understanding the phenomenology
of neutrino oscillations, $Z$-strahlung, pion decay kinematics 
and the consistency of this result with
supernova 1987A data.

\end{abstract}

\maketitle

{\bf Introduction}\\[-.1in]

Lorentz invariance has been foundational in formulating both the Standard Model of particle physics and general relativity.  Remarkably precise experimental tests have been carried out to check its validity.  One generally quantifies the accuracy of such tests by adding small Lorentz-invariance-violating terms to a conventional Lagrangian, with data from   experiments subsequently used to set  upper bounds on the coefficients of these terms.  For example modifying the square of the magnetic field relative to all other terms in the  Lagrangian of quantum electrodynamics yields a  photon
velocity $c_\gamma$ that differs from the maximum attainable 
velocity of a material body $c_M$
\cite{HW}.  This perturbation breaks 
Lorentz invariance, preserving rotational and translational invariance in one frame (the preferred frame) but not in any other frame. It is common to interpret the preferred frame as that for which the cosmic microwave background is isotropic, in which case small anisotropies  appear in laboratory experiments.  High-precision spectroscopic 
experiments have set a bound of    $|1- c^2_\gamma/c^2_M | < 6 \times 10^{-22} $ \cite{photobound}.
 
A Lorentz transformation of the coordinates for an observer moving at velocity $\vec v$   yields
\begin{equation}\label{eq2}
 \vec p = \gamma ~ m ~\vec v\,; ~~~ {\rm and} ~~~ E = \gamma ~ m ~ c^2\,,
\end{equation}
for the energy and momentum  of a body, where $\gamma = 1/\sqrt{1 - v^2/c^2}$ is the Lorentz factor. 
The invariant momentum is defined as
\begin{equation}\label{eq3}
 p^2 = p_\mu p^\mu = \left(\frac{E}{c}\right)^2 - |\vec p|^2 = m^2 c^2 \,,
\end{equation}
so that 
\begin{equation}\label{eq4}
 E^2 = |\vec p|^2  c^2 + m^2 c^4 \,.
\end{equation}
For any massive particle we thus require $v < c$, otherwise the Lorentz
factor $\gamma$ becomes imaginary and the total energy cannot be
positive definite. This contradicts the recent measurement of the muon-neutrino velocity with the OPERA detector at Gran Sasso of the CNGS
$\nu_\mu$ beam from CERN \cite{Opera}. Furthermore, it is straightforward to show
(using equation \ref{eq2})
that  for a muon neutrino with mass around $m \sim 1$ eV and 
energy $E \sim 17$ GeV that $(c - v )/c \sim 10^{-20}$. 

Should this result be confirmed by subsequent experiments, it will require a significant revision
of the Standard Model.  Constructing such a revision will be a major challenge, since any new model
must be consistent with astrophysical bounds (from SN1987A \cite{sn})  and neutrino oscillation experiments,
as well as explain the OPERA data without undermining basic notions of cause and effect.

\vskip 2mm
{\bf Violation of Lorentz Invariance}\\[-.1in]

We argue here that the present measurement of the neutrino velocity 
could be consistent with causality if Lorentz
invariance is violated for the interactions of the muon neutrinos.  
We regard this as the most conservative interpretation of the
OPERA data in conjunction with other experiments.  It retains the
essential foundations of relativity, namely that the velocity of light (photons)
 is constant and the causal structure of spacetime is preserved.  The validity of
 Lorentz invariance beyond the first generation of the Standard Model has never before been directly 
 tested, and so its violation is not empirically ruled out \cite{Hambye:1998sm, Hambye:1997jy,koste}. 
 Violation  of   Lorentz invariance (VLI) has also been 
 considered to evade the GZK cosmic ray cut off and explain the ultra high energy cosmic
 rays \cite{GZK}, taking the strong constraints \cite{CG,CR} 
 on the parameters coming from studies of cosmic rays into consideration.  
Lorentz invariance violation has also been studied extensively in extensions of the
 standard model and constraints on various model parameters have been reviewed \cite{Kos}.

We follow the analysis of Coleman and Glashow \cite{CG}, where they made a
general construction of Lorentz non-invariant interactions of 
particles, keeping the gauge invariance intact. 
We introduce a Lorentz non-invariant interaction term for the
muon neutrinos 
\begin{equation}\label{eq5}
 i ~u^\dagger ~ \left[ D_0 - i \left( 1 - \frac{1}{2} \epsilon
 \right) \vec \sigma \cdot \vec D \right] ~u \,.
\end{equation}
The parameter $\epsilon$ is a measure of violation of
Lorentz invariance. This will modify the high energy behaviour
of the muon neutrinos and modify its energy momentum relation.
The renormalized Lorentz invariant propagator 
\begin{equation}\label{eq6}
 i ~S_F(p) = \frac{i}{ (\psl - m) A(p^2)} \,,
\end{equation}
for some function $A(p^2)$ normalized as $A(m^2)=1$, 
will be modified to 
\begin{equation}\label{eq7}
 i ~S_F(p) = \frac{i }{( \psl - m) A(p^2) 
 + \frac{1}{2} \epsilon ~\vec \gamma \cdot \vec p ~ B(p^2)} \,,
\end{equation}
where we normalize the Lorentz non-invariant interaction as
$B(m^2)=1$. 

The lowest order shift in the poles of the propagator then 
gives
\begin{equation}
 p^2 =  E^2 - |\vec p|^2 = m^2 + \epsilon |\vec p|^2\,.
\end{equation}
This implies that the Lorentz non-invariant contribution added
a shift in the momentum, which will result in a shift in the 
maximum attainable velocity of the particle from the velocity
of light $c$ (for our choice of units $c = 1$, which remains the
maximum attainable velocity for all other particles except the
muon neutrino) to a new value  
\begin{equation}
 c^2_{\nu_\mu} = (1 + \epsilon) c^2   = (1 + \epsilon) 
\end{equation}
The energy momentum relation then becomes
\begin{equation}
 E^2 = |\vec p|^2 c_{\nu_\mu}^2 + m_{\nu_\mu}^2 c_{\nu_\mu}^4 \,.
\end{equation}
The muon neutrino mass has now shifted by a factor, $m_{\nu_\mu} =
m/(1 + \epsilon)$, so that we still have $m_{\nu_\mu} c_{\nu_\mu}^2 = m $
to be the rest mass of the muon neutrino. 
The Lorentz factor will also be shifted because of this Lorentz
non-invariance, and the new Lorentz factor is given by
\begin{equation}
 \gamma_{\nu_\mu} = \frac{1}{ \sqrt{1 - v^2/c_{\nu_\mu}^2}} \,,
\end{equation}
so that the mass energy relation becomes
\begin{equation}
 E = \gamma_{\nu_\mu} m_{\nu_\mu} c_{\nu_\mu}^2 = \gamma_{\nu_\mu} m 
\end{equation}
and
\begin{equation}
\frac{(c_{\nu_\mu} - v)}{c_{\nu_\mu}} \approx 
 \frac{m^2}{2 E^2} \,, 
\end{equation}
for very large $v$.

These modifications will now allow us to interpret the recent
measurement of the neutrino velocity $v$ from the OPERA detector.
The present result from OPERA gives
\begin{equation}
  \frac{(v - c)}{c} = v-1 =  (2.48 \pm 0.28 \pm 0.3) \times
 10^{-5} \,.
\end{equation}
However, in the presence of the Lorentz invariance
violating interactions, the Lorentz factor that the
muon neutrinos experience contains the factor
\begin{equation}\label{result}
 \frac{(c_{\nu_\mu} - v)}{c_{\nu_\mu}} \approx \frac{m^2}{2 E^2} 
 \approx 10^{-20} \,.
\end{equation}
This determines the amount of violation of Lorentz invariance,
given by
\begin{eqnarray}
 \epsilon &=& \frac{c^2_{\nu_\mu} }{c^2}- 1 =
 \left[\frac{(c_{\nu_\mu} - v)}{c} + 
 \frac{( v - c)}{c}   \right]\left(\frac{c_{\nu_\mu}}{c}+1\right) \nonumber\\
 & \approx& 2 \left[\frac{(c_{\nu_\mu} - v)}{c_{\nu_\mu}} + 
 v-1  \right]
  \end{eqnarray}
Since $v < c_{\nu_\mu}$, the 
present measurement of the muon neutrino velocity $v$ (with $v>c$) becomes consistent
with the modified Lorentz factor and the muon neutrino remains time-like 
in the same context.

\vskip 2mm
{\bf Models for VLI}\\[-.1in]
 
In a recent article it has been pointed out that the models
of VLI suffer from a serious constraint coming from the pion
lifetime kinematics \cite{cn}. The amount of VLI required to explain the
OPERA result is in contradiction with present data from accelerators
and cosmic rays. We thus propose here a couple of scenarios that
may give rise to an effective VLI for  muon neutrinos that
does not affect the pion lifetime. These models are based on the
fact that the effective VLI originates from an interaction of
the propagating neutrinos with the  environment, so during 
a pion decay there is no VLI.

In the first solution we consider the possibility that the neutrinos
interact with the background dark energy, which gives rise to an
effective VLI. We consider the scenarios, in which neutrinos form 
condensates after they acquire masses and that explains the observed
dark energy at present times \cite{de}. This background neutrino 
condensate dark energy can, in principle, affect the dynamics of the 
neutrinos, compared to other particles. For example, a 
$\nu_\mu$ with momentum $p$ can collide with a condensate 
$ \overline{\nu}_a - \nu_\mu$ pair and bind with the $ \overline{\nu}_a$.  The liberated
$\nu_\mu$, located at a distance $x$ away from its condensate partner,
will continue with momentum $p$ due to momentum conservation.  As this process is repeated,
the net effect is that the $\nu_\mu$ ``hops" through the condensate at an effective speed
greater than unity, resulting in a different maximum
attainable velocity for the muon neutrinos.
Only the maximum attainable velocity of the neutrinos can
be affected by this mechanism.  This yields an interesting solution to the OPERA result, because the velocity of the muon neutrino becomes
larger than the velocity of light when the muon neutrinos interact with the
background dark energy.  It evades constaints due to $Z$-strahlung radiation 
($\nu_\mu \to \nu_\mu + Z \to \nu_\mu +e^+  + e^-$)  
\cite{CohG}  since the $\nu_\mu$ does not actually travel faster than the speed of light 
as it moves through the condensate.  The strong constraints coming from  observed pion decays \cite{cn}
and cosmic rays are also not applicable, whereas these constraints
can cripple many of the conventional models of VLI.

We now turn to the second solution for an effective VLI,
which is through a violation of the equivalence principle (VEP),
considered previously as a mechanism
for neutrino oscillations \cite{gnuosc}. An equivalence of
this type of VEP and an effective VLI at the phenomenological 
level has already been demonstrated \cite{halp}, although
their origin is completely different. 
The violation of equivalence principle will
introduce a shift in the momentum in a constant gravitational potential
if the neutrinos couple to gravity with a different coupling,
which can be treated as equivalent to a shift in the maximum attainable velocity of the 
neutrinos. This mechanism satisfies the $Z$-strahlung constraints \cite{CohG}
because  neutrinos effectively follow subluminal geodesics of a different metric
than other particles due to VEP, and so do not undergo the intense $Z$-strahlung effect
(see figure 1).
This proposal is also unaffected by the pion decay  constraint, because the neutrino
velocity becomes large only when it propagates in a background
gravitational field.

Without going into the details of the mechanism that gives us the violation
of equivalence principle, we just assume that a background gravitational 
potential $\phi$ 
changes the metric $g^{44} =  (1 + \alpha \phi) $, which in turn gives
a correction to the energy of the test neutrinos
\begin{equation}
 E_0 = p + \frac{m^2}{2 p} - \alpha \phi p \,,
\end{equation} 
where $\alpha$ could be different for different materials, in
violation of the weak equivalence principle, and is a measure
of the amount of VEP. From this correction it is clear that
$\alpha \phi c$ may be considered as the change in the
maximum attanable velocity of the neutrinos giving rise to an
effective VLI. 
 
\vskip 2mm
{\bf Neutrino Phenomenology}\\[-.1in]

Turning to the phenomenology of our proposal, the simplest scenario to explain neutrino
oscillation data with an effective Lorentz invariance violation (EVLI)
is to assume that
all the three neutrinos have the same maximum attainable velocity, 
$c_{\nu_e} = c_{\nu_\mu} = c_{\nu_\tau} = c_\nu  \approx (1 + 2.5 \times 10^{-5})c$
and there is no energy dependence. These assumptions are  consistent with
MINOS \cite{minos}, the short baseline experiments \cite{mino1},
and also the present observation at OPERA in the two energy bins of 13 GeV
and 42.9 GeV, but fail to explain the supernovae
SN1987A bound \cite{sn} of  $|1-c_\nu/c| < 10^{-9} $,  
where $\bar{\nu_e}$ was observed. If we further assume that the there is
no EVLI for any other particles, there are no other phenomenological constraints.
We shall now consider a few possibilities
to explain the supernovae bound along with the neutrino oscillation
constraints.

\begin{itemize}
\item We consider that the neutrino condensates changes the 
maximum attainable velocities of all the three neutrinos by same
amount, so there is no constraint from neutrino oscillation
experiments. However, during their propagation through intergalactic
medium, they experience a new source of VEP, which slows them down
and they satisfy the supernovae bound.
\item The VEP in the dark matter background changes because they interact
in a different manner compared to interaction of VEP with
ordinary matter. During the propagation of the neutrinos on Earth,
they experience the effect of VEP in the background of ordinary
matter, which gives an EVLI that can explain the OPERA result. But
during the propagation of the neutrinos from supernovae, they 
experience VEP in the dark matter background, which makes the
neutrinos slower explaining the supernovae bound. 
\item This solution is similar to that of reference \cite{hann},
where a light sterile neutrino is introduced, but the model is
similar to that of reference \cite{de}.
We assume the existence of four neutrino states, 
$\nu_1, \ldots, \nu_4$, three of
which have the same maximum attainable velocity similar to other particles 
$c_{\nu_1} = c_{\nu_2} = c_{\nu_3} = c =1$, with the 
fourth having $c_{\nu_4} \approx (1 + 2.5 \times 10^{-5})c$,
because the $\nu_4$ condensates explains the dark energy.  
 This fourth  neutrino  is a combination of $\nu_\mu$ and a 
sterile neutrino $\nu_s$:
$ \nu_4 = \nu_\mu + \nu_s^c; $ and  has 
negligible mixing with $\nu_e$. After production, muon neutrinos will propagate
partly as the fourth physical state $\nu_4$ 
because its wave function has a large component of $\nu_4$. 
As a result the velocity of the muon neutrino will appear to be 
similar to $c_{\nu_4}$, explaining the
observation at OPERA and MINOS.  However 
neutrinos emitted from supernova SN1987A travel mostly as $\nu_1$, 
with only a small fraction traveling
as $\nu_4$.  Consequently  the required constraints from the SN1987A data are respected.
\end{itemize}

\begin{figure}
\includegraphics[width=5cm]{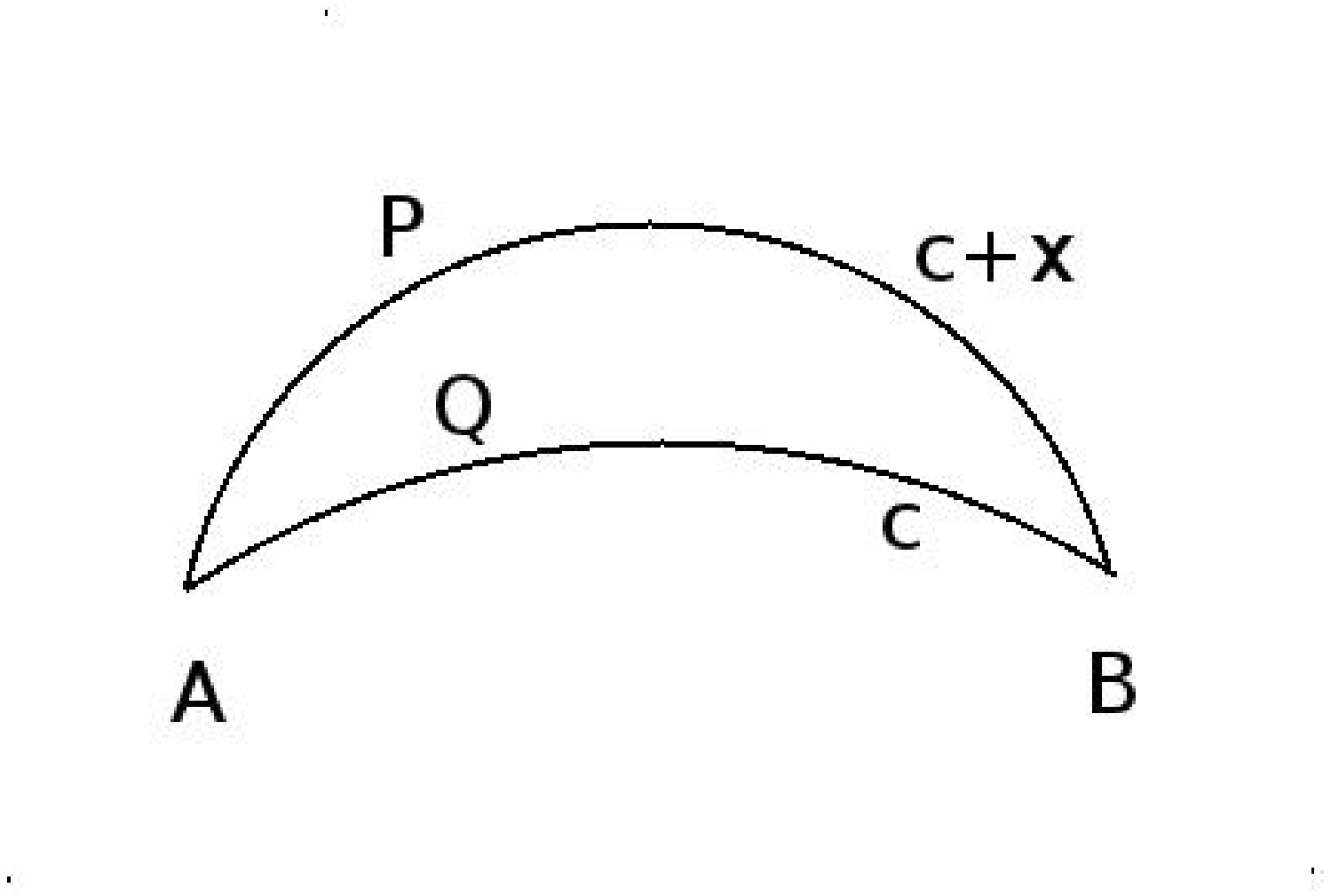}  
\caption{\label{vep} P is the geodesic of all particles, except neutrinos,
while Q is the geodesic of neutrinos. So, light travels a distance $c+x$ between
A B, while $\nu_\mu$ travels a distance $c$. }
\end{figure}

\vskip 2mm
{\bf Comments}\\[-.1in]

It is crucial to repeat the OPERA experiment. Should its findings be confirmed
its implications for the foundation of physics will be profound.  We have proposed what
is perhaps the minimal change in this regard:  VLI or VEP implemented in such a way that
all neutrino species have the same effective maximal attainable velocity. From this perspective
the photon  moves at a speed slower than this maximum.  However this cannot be 
due to a slightly massive photon.  Current upper bounds on the photon mass range from $10^{-14}$ eV to $10^{-24}$ eV
\cite{GN}.  Taking the more stringent bound of $m_\gamma \sim 10^{-24}$, then from $E= \frac{ m c_{max}^2}{\sqrt{1-v_\gamma^2/c^2_{max} } }$, we have from $E\sim 17$ GeV, $v_\gamma - c_{\max} \sim 10^{-68}$, so $c_{\nu_\mu} \lesssim c_{\max}$ cannot explain  the OPERA result.  The slower photon speed could be due to an aether whose
motion is so tightly correlated with that of the earth that the stringent anisotropy constraints of $c_{\theta}/c < 10^{-15}$ are evaded \cite{Muller}. 
This is not unreasonable if
the aether is the condensate dark energy mentioned above,  since (apart from neutrinos) the earth is therefore composed of particles that respond to the aether exactly as the photon does.  Yet another possibility is that the universe is governed by a bi-metric theory, with all particles coupling to  one metric except for neutrinos, which couple to the other metric
whose null rays trace out light cones with slope $c > 1$.  
Neutrinos  emitted at
high-enough energy from  active galactic nuclei, supernovae, or some other astrophysical phenomenon, will then yield  Cerenkov-type effects that may be observable. 
 
\vskip 2mm
{\bf Summary}\\[-.1in]

To summarize, we have pointed out that the OPERA data can be consistent 
with all other experiments if there 
there is a maximal speed larger than the speed of light that only  neutrinos can approach.  
The models we have proposed satisfy constraints from neutrino oscillation, SN1987A, and pion decay kinematics, Z-Strahlung,
and are consistent with OPERA/MINOS data.
Whether or not this class of models survives future empirical scrutiny remains to be seen.

\vskip 2mm
{\bf Note Added} Many interesting works followed the announcement of the OPERA
result, we mention here a few of them \cite{opx}.

\vskip 2mm
{\bf Acknowledgements}
This work was supported in part by the Natural Sciences and Engineering 
Research Council of Canada. U. Sarkar would like to thank Prof. R. Cowsik, Director, McDonnell Center for the Space Sciences, Washington University in St. Louis, 
for arranging his visit as the Clark Way Harrison visiting professor.


\begin{thebibliography}{99}

\bibitem{HW} M.P. Hagen and C.M. Will, Phys. Today {\bf 40}, 69 (1987);
Y. Grossman, C. Kilic, J. Thaler, and D.G.E. Walker, Phys. Rev. {\bf D 72}, 125001 (2005).

\bibitem{photobound}S. K. Lamoreaux, J. P. Jacobs, B. R. Heckel, F. J. Raab, 
and E. N. Fortson, Phys. Rev. Lett. {\bf 57}, 3125 (1986); 
C.J. Berglund, L.R. Hunter, D. Krause, Jr., E.O. Prigge, M.S. Ronfeldt, and 
S.K. Lamoreaux,  Phys. Rev. Lett. {\bf 75}, 1879 (1995).

\bibitem{Opera} {\it OPERA Collaboration}: T. Adam, et al, arXiv:1109.4897v1 [hep-ex].

\bibitem{sn} K. Hirata, et al, Phys. Rev. Lett. {\bf 58}< 1490 (1987);
R.M. Bionta, et al, Phys. Rev. Lett. {\bf 58}, 1494 (1987);
M.J. Longo, Phys. Rev. {\bf D 36}, 3276 (1987).

\bibitem{Hambye:1998sm}
  T.~Hambye, R.~B.~Mann, U.~Sarkar,
  Phys.\ Rev.\  {\bf D58}, 025003 (1998).
  [hep-ph/9804307].

\bibitem{Hambye:1997jy}
  T.~Hambye, R.~B.~Mann, U.~Sarkar,
  Phys.\ Lett.\  {\bf B421}, 105 (1998).
  [hep-ph/9709350].
  
\bibitem{koste} V.A. Kostelecky, Phys. Rev. Lett. {\bf 80}, 1818 (1998);
Phys. Rev. {\bf D 61}, 016002 (1999); 	
R. Bluhm, V.A. Kostelecky, and C.D. Lane, Phys. Rev. Lett. {\bf 84}, 1098 (2000). 

\bibitem{GZK} S. Coleman and S.L. Glashow, hep-ph/9808446 (1998);
F.W. Stecker and
S.T. Scully, Astroparticle Physics {\bf 23}, 203 (2005);
L. Maccione, A.M. Taylor, D.M. Mattingly, and S. Liberati, JCAP {\bf 0904}, 022 (2009).

\bibitem{CG} S. Coleman and S.L. Glashow, Phys. Rev. {\bf D 59}, 
116008 (1999).

\bibitem{CR} S. Coleman and S.L. Glashow, Phys. Lett. {\bf B 405}, 249 (1997);
R. Cowsik and B.V. Sreekantan, Phys. Lett. {\bf B 449}, 219 (1999);
X.-J. Bi, P.-F. Yin, Z.-H. Yu, and Q. Yuan, arXiv:1109.6667 [hep-ph] (2011).

\bibitem{Kos} V.A. Kostelecky and N. Russel, Rev. Mod. Phys. {\bf 83}, 11 (2011). 

\bibitem{CohG}A. Cohen and S. Glashow, arXiv: 1109.6562 [hep-ph] (2011).

\bibitem{cn} R. Cowsik, S. Nussinov and U. Sarkar, arXiv:1110.0241 [hep-ph] (2011).

\bibitem{de} J.R. Bhatt, B. Desai, E. Ma, G. Rajasekaran and U. Sarkar,
Phys. Lett. {\bf B 687}, 75 (2010).

\bibitem{gnuosc} M. Gasperini, Phys. Rev. {\bf D 38}, 2635  (1988); 
M. Gasperini, Phys. Rev. {\bf D 39},  3606 (1989);
J.T. Pantaleone, A. Halprin, and C.N. Leung, Phys. Rev. {\bf D 47}, 4199 (1993) [hep-ph/9211214]; 
J.R. Mureika and R.B. Mann, Phys.Rev. {\bf D54}, 2761 (1996); 
R.~B.~Mann, U.~Sarkar,   Phys.\ Rev.\ Lett.\  {\bf 76}, 865 (1996).
  [hep-ph/9505353].

\bibitem{halp} A. Halprin and H.B. Kim, Phys. Lett. {\bf B 469}, 78 (1999).


\bibitem{minos}
{\it MINOS Collaboration}: P. Adamson, et al, Phys. Rev. {\bf D 76}, 072005 (2007).

\bibitem{mino1} G.R. Kalbfleisch, N. Baggett, E.C. Fowler, J. Alspector, 
Phys. Rev. Lett. {\bf 43}, 1361 (1979);
J. Alspector, et al, Phys. Rev. Lett. {\bf 36}, 837 (1976).

\bibitem{hann} S. Hannestad and M.S. Sloth, arXiv:1109.6282 [hep-ph] (2011); 
I.Ya. Aref'eva, and I.V. Volovich, arXiv:1110.0456 [hep-ph] (2011);
H. Pas, S. Pakvasa, T.J. Weiler, Phys. Rev. {\bf D 72}, 095017 (2005).


\bibitem{GN}
 A.~S.~Goldhaber, M.~M.~Nieto,
  Rev.\ Mod.\ Phys.\  {\bf 82}, 939  (2010).
  [arXiv:0809.1003 [hep-ph]].
  
\bibitem{Muller} H. Muller, S. Herrmann, C. Braxmaier, S. Schiller, and A. Peters, 
Phys. Rev. Lett. {\bf 91} 020401 (2003).
  
\bibitem{opx}  J. Alexandre, J. Ellis and N.E. Mavromatos, arXiv:1109.6296 [hep-ph] (2011);
G.F. Giudice, S. Sibiryakov, and A. Strumia, arXiv:1109.5682 [hep-ph]
(2011); R.~A.~Konoplya, arXiv:1109.6215 [hep-th] (2011);
 C.~Pfeifer, M.~N.~R.~Wohlfarth, arXiv:1109.6005 [gr-qc] (2011);
N.D. Haridass, arXiv:1110.0351 [hep-ph] (2011); T. Li and D.V. Nanopoulos, 
arXiv:1110.0451 [hep-ph] (2011); H. Gilles, arXiv:1110.0239 [hep-ph] (2011); 
G. Amelino-Camelia, G. Gubitosi, N. Loret, F. Mercati, G. Rosati,
and P. Lipari, arXiv:1109.5172 [hep-ph] (2011);
P. Wang, H. Wu, and H. Yang, arXiv:1110.0449 [hep-ph] (2011);
arXiv:1109.6930 [hep-ph] (2011).

\end{thebibliography}
\end{document}